\pdfoutput=1

\documentclass[11pt,twoside,a4paper,cmspaper,final,collab]{cms-tdr}
\def\svnVersion{457142P}\def\cmsCernNoTag{CERN-EP-2017-124}\def\cmsCernDate{\today}\def\cmsMessage{Published in the Journal of High Energy Physics as \href{http://dx.doi.org/10.1007/JHEP03(2018)167}{\doi{10.1007/JHEP03(2018)167.}}}

\begin{document}\cmsNoteHeader{SUS-15-009}

\hyphenation{had-ron-i-za-tion}
\hyphenation{cal-or-i-me-ter}
\hyphenation{de-vices}
\RCS$Revision: 447940 $
\RCS$HeadURL: svn+ssh://svn.cern.ch/reps/tdr2/papers/SUS-15-009/trunk/SUS-15-009.tex $
\RCS$Id: SUS-15-009.tex 447940 2018-02-25 02:27:12Z alverson $
\providecommand{\SDECAY}{\textsc{sdecay}\xspace}

\title{Search for natural supersymmetry in events with top quark pairs and photons in pp collisions at $\sqrt{s}=8\TeV$}

\date{\today}

\abstract{
Results are presented from a search for natural gauge-mediated supersymmetry (SUSY) in a scenario
in which the top squark is the lightest squark, the next-to-lightest SUSY particle is a bino-like
neutralino, and the lightest SUSY particle is the gravitino.  The strong
production of top squark pairs can produce events
with pairs of top quarks and neutralinos, with each bino-like neutralino
decaying to a photon and a gravitino. The search is performed using a sample of
pp collision data accumulated by the CMS experiment at $\sqrt{s}=8\TeV$, corresponding to an integrated
luminosity of 19.7\fbinv. The final state consists of a lepton (electron or muon), jets,
and one or two photons. The imbalance in transverse momentum in the events is compared with the expected
spectrum from standard model processes. No excess event yield is observed
beyond the expected background, and the result is
interpreted in the context of a general model of gauge-mediated
SUSY breaking that leads to exclusion of top squark masses below 650--730\GeV.}

\hypersetup{%
pdfauthor={CMS Collaboration},%
pdftitle={Search for natural supersymmetry in events with top quark pairs and photons in pp collisions at sqrt(s) = 8 TeV},%
pdfsubject={CMS},%
pdfkeywords={CMS, physics, top quark, photon, SUSY}}

\maketitle

\section{Introduction}
\label{intro}

Supersymmetry (SUSY)~\cite{Wess,Golfand:1971iw,Volkov:1972jx,Chamseddine,Kane,Fayet,Barbieri,Hall,Ramond}
is an extension of the standard model (SM) that can avoid
the hierarchy~\cite{tHooft:1979rat,Witten,Dine100,Raby,Georgi,Kaul} problem confirmed by
the recent discovery~\cite{Aad:2012tfa,Chatrchyan:2012ufa} of the Higgs boson at a mass of
125\GeV~\cite{Aad:2015zhl}. Of
particular interest are the ``natural'' models of SUSY in which several SUSY partners (sparticles)
of SM particles are expected to be light,
such as the third-generation squarks and the Higgsinos~\cite{Barnard:2012,Kats:2011}.
In such SUSY scenarios, the fine-tuning required to maintain low mass of the electroweak scale can be avoided.
Many light top squark searches are being
pursued at the LHC~\cite{Sirunyan:2016jpr,Chatrchyan:2013mya,Khachatryan:2014doa,Aad:2012jva,Aad:2015hea,Aaboud:2016lwz}.

\begin{figure}[tbph]
  \centering
    \includegraphics[width=0.6\textwidth]{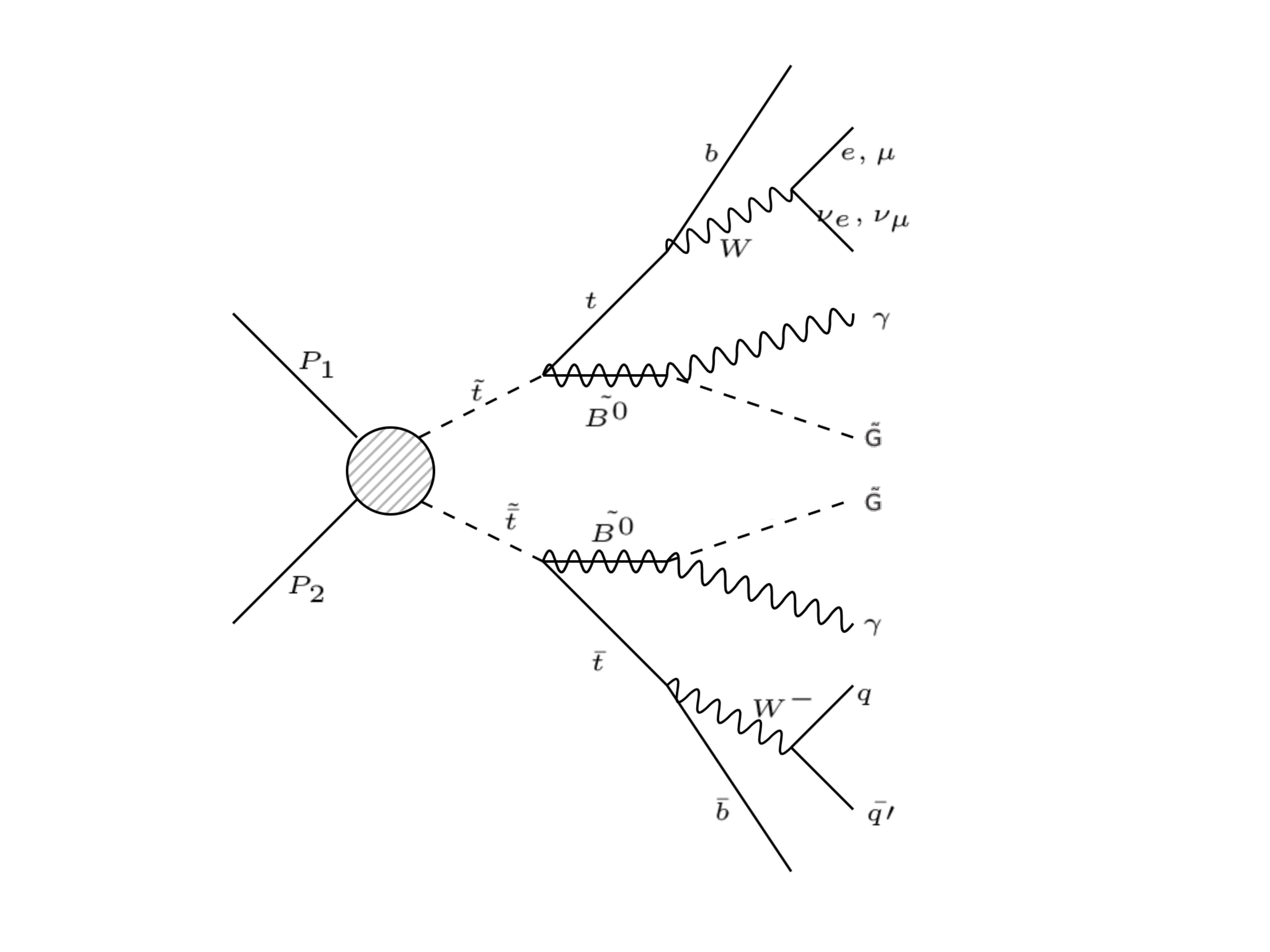}
    \caption{The event topology used to search for low mass top squarks pairs
}
    \label{fig:feynman_diagram}

\end{figure}

In this paper, we describe a search for light top squarks (\sTop)
in a data sample corresponding to an integrated luminosity of 19.7\fbinv of pp collisions at
$\sqrt{s}=8\TeV$. This search is motivated by models of gauge-mediated
SUSY  breaking (GMSB)~\cite{Dine:1981gu,Nappi:1982hm,AlvarezGaume:1981wy}
in which the neutralino ($\chiz_1$) is the next-to-lightest sparticle (NLSP) and the gravitino ($\tilde{G}$)
is the lightest sparticle (LSP). The gravitino escapes undetected
and contributes to missing transverse momentum (\ptvecmiss) in the
detector, where the magnitude of \ptvecmiss is referred to as
\ptmiss. This search considers a bino-like neutralino that decays
to a photon and a gravitino.
Assuming that $R$-parity~\cite{Barbier:2004ez,rparity} is conserved, pair production
of sparticles would be the dominant production
mechanism for SUSY particles in pp collisions at the LHC.
Because top squarks are expected to be relatively light in natural SUSY scenarios, we search for top squark
pair production, a strong process.
Assuming a bino-like neutralino NLSP, each top squark would decay
to a top quark and a neutralino, with the  neutralino decaying
to a photon and a gravitino, leading to a $\ttbar$+\Pgg\Pgg+\ptmiss topology. This event topology is shown
in Fig.~\ref{fig:feynman_diagram}.

The analysis concentrates on the semileptonic decay of the \ttbar
pair, thereby requiring the presence of exactly one
isolated electron or  muon.  This minimizes contributions from multijet
and \Pgg+jet backgrounds. At least one jet in each event is required to be tagged
as originating from a b quark to reduce non-\ttbar backgrounds. No
explicit $\ttbar$+\Pgg\Pgg~sample is used in the background estimates because of
the exceedingly small cross section for such events in the SM.
Two signal regions are defined for both electron and
muon channels, depending on the presence of one or two selected photons in the
event. Control regions are similarly defined, using photons that fail either
the nominal isolation or shower-energy distribution requirements.

The results of the analysis are evaluated by comparing the shapes of \ptmiss
distributions between the data and estimated backgrounds in the one- and
two-photon signal regions.  The results are interpreted for  a
range of top squark and neutralino masses in a general gauge-mediated (GGM) SUSY model
framework~\cite{Meade:2008wd,Buican:2008ws,Abel:2009ve,Ruderman:2011vv,Kats:2012ym,Grajek:2013ola}.

\section{The CMS detector}
\label{detector}

The central feature of the CMS detector is a superconducting solenoid
with an internal diameter of 6\unit{m}, providing a magnetic field of 3.8\unit{T}. A silicon pixel and strip tracker,
a lead tungstate crystal electromagnetic calorimeter (ECAL), and a brass and scintillator sampling hadron calorimeter
(HCAL), each separated into central barrel and endcap sections, reside within the field volume. Extensive
forward calorimetry complements the coverage provided by the barrel and endcap detectors. The muon system,
embedded in the steel return
yoke outside of the solenoid, measures muons using drift tubes, cathode strip chambers, and resistive plate chambers.

An energy resolution averaging approximately 1\% is achieved for unconverted or late-converting photons in
the energy range of photons in the barrel section of ECAL.
The remaining converted barrel photons have a resolution of about 1.3\% up to a pseudorapidity of $\abs{\eta} = 1$,
rising to about 2.5\% at $\abs{\eta} = 1.4$~\cite{Khachatryan:2015iwa}. Only photons located in the barrel of the
ECAL are considered in this analysis because of the
superior energy resolution in the barrel compared to the endcap.

The first level of the CMS trigger system, constructed using special hardware processors,
provides information from the
calorimeters and muon detectors to select the most interesting events in a fixed time interval of less than 4\mus.
The high-level trigger processor farm further decreases the event rate from around 100\unit{kHz} to about
400\unit{Hz} before data storage.

A more detailed description of the CMS detector, together with a definition of the coordinate system and the
kinematic variables, such as $\eta$ or the
azimuthal angle $\phi$ (in radians), can be found in Ref.~\cite{Chatrchyan:2008zzk}.

\section{Object reconstruction}
\label{reconstruction}

All physics objects in the event (muons, electrons, photons, jets, and \ptmiss)
are reconstructed using the particle-flow (PF)
algorithm~\cite{CMS-PAS-PFT-09-001,CMS-PAS-PFT-10-001}. Jets
are formed by clustering PF candidates using the anti-\kt
algorithm~\cite{Cacciari:2008gp}, as implemented in \FASTJET toolkit~\cite{fastjet}, using
a distance parameter of 0.5, and their
momenta are corrected for effects of multiple interactions in the
same or neighboring bunch crossings (pileup). The \ptmiss of an event is defined
by the projection of the negative of the vector sum of the
momenta of all reconstructed objects in the event onto
the plane perpendicular to the proton beams.
All PF candidates are used in the calculation of \ptmiss.

Photons are reconstructed from energy clusters in the ECAL barrel
($\abs{\eta} < 1.44$), are required to be highly isolated from other objects,
and to have transverse momentum  $\pt>20\GeV$. The ratio of the energy deposited in the HCAL tower closest
to the seed of the ECAL photon cluster to
the energy in the photon cluster has to be less than five percent.
The photon shower is required to have a photon-like spatial distribution
in its energy~\cite{Khachatryan:2015iwa}. The isolation variable, defined through
the sum of the scalar values of $\pt$ of all PF candidates within a cone centered on the photon axis,
in the $\eta$-$\phi$ plane of $\Delta R = \sqrt{\smash[b]{(\Delta\phi)^2 + (\Delta\eta)^2}}=0.3$,
is calculated without including the $\pt$ of the candidate photon.
The isolation energy for charged hadrons is required to be $<$15\GeV, the neutral-hadron energy
$<$3.5\GeV +  4\% of the photon candidate $\pt$, and the isolation energy from any other photons in the cone must be
$<$13\GeV + 0.5\% of the candidate photon $\pt$.
Pileup corrections depending on $\eta$ are applied to all calculated isolation variables.

Electrons are reconstructed from clusters of deposited energy in the ECAL that are matched to a track in the
silicon tracker~\cite{cmsElectronReco}. Candidate electrons are
required to have $\pt>30$\GeV, and to be within $\abs{\eta} < 2.5$, excluding
the small transition region ($1.44<\abs{\eta} < 1.52$) between the ECAL barrel and the endcaps.  Electrons
are required to be isolated, with the sum of the energy deposition within a
cone of radius $\Delta R = 0.3$, excluding the electron,
to correspond to $<10\%$ of the momentum of the candidate electron.

Muons are reconstructed from measurements in the muon system and
compatible track segments in the silicon tracker~\cite{Chatrchyan:2012xi}.
Candidate muons are required to have
$\pt > 30$\GeV, be within $\abs{\eta} < 2.1$, and to have an isolation
energy sum in a cone of radius $\Delta R = 0.4$, excluding the muon, of $<$12\% of
their $\pt$.  Looser lepton requirements are applied to identify extra leptons that are
used to veto the dilepton \ttbar final states, as described in Section 4.

The combined secondary vertex algorithm (CSV)~\cite{CMS-PAS-BTV-12-001,CMS-PAS-BTV-13-001} is used to
identify jets from b quarks. The CSV algorithm uses
secondary vertices and track impact parameters to provide a
discriminant separating b quark jets from charm, light quark, or gluon
jets. The selection efficiency is about 70\% for b quark jets and
20\% for c quark jets. The rejection factor for lighter quark or
gluon jets at this working point is about 2\%.

\section{Event selection and analysis strategy}
\label{analysis}

Events are required to pass either a single-electron or single-muon
trigger, requiring one isolated electron or muon with minimum
$\pt$ of 27 or 24\GeV, respectively. In addition, the single-muon trigger
requires the muon candidate to be within $\abs{\eta} < 2.1$. The trigger efficiency
is approximately 100\% using offline cuts on $\pt$ of 30\GeV.

Only one lepton and at least three jets with $\pt >
30$\GeV and $\abs{\eta} < 2.4$ are required, with at least one of the three jets
tagged as originating from a b quark. All objects are
required to be separated from each other by at least $\Delta R = 0.5$.
Events containing additional leptons satisfying less restrictive criteria of $\pt > 10$\GeV,
$\abs{\eta}<2.5$, and isolation-energy sums with $<$20\% of their
$\pt$, are rejected.

After this preselection, events are separated into independent samples
based on the number of candidate photons. Candidate photons
are required to be separated from all jets by $\Delta R > 0.7$.
Two signal regions are defined, with SR1 containing
one photon candidate, and SR2 at least two photon
candidates.

Photons that fail either the shower-energy distribution or charged-hadron isolation
criteria are referred to as fake
photons. These objects are predominantly jets with large electromagnetic
fluctuations in their hadronization and are used to define two control regions:
CR1, containing one fake and no properly reconstructed photons, and CR2, containing
two or more fake and no properly reconstructed  photons.
The control regions are defined not to overlap with
signal regions, to have very small acceptance for signal, and to greatly
enhance the population of photon-like jets that contribute most
of the estimated background in signal regions. The control regions
also provide events that can be used to study  the performance of the \ptmiss  simulation
for  poorly reconstructed photon-like objects in the signal region.
The effect on the \ptmiss  resolution from these poorly reconstructed photon-like objects
is found to be negligible compared to the effect of $\pt$ resolutions in the
jets from the \ttbar decays.

The background expected in the signal regions is largely dominated by
$\ttbar$+jets and $\ttbar$+\Pgg~events, where many selected photons may originate from
misreconstructed jets. These two processes are simulated in Monte Carlo
(MC) using the leading-order (LO) \MADGRAPH~5.1.3~\cite{Alwall:2011uj}
matrix element generator matched to \PYTHIA~6.426~\cite{Sjostrand:2006za}
for parton showering and fragmentation. Simulated $\ttbar$+\Pgg~ events are generated in a
$2 \to 7$ configuration (pp$\to$bbjj$\ell\nu$\Pgg). Approximately 0.6\% of the simulated
$\ttbar$+jets events that contain a generator-level photon fall into the phase
space of the $\ttbar$+\Pgg~sample, and are removed to avoid double counting these events.
Most other backgrounds
are simulated with \MADGRAPH and matched to \PYTHIA, including
W+jets or Z+jets, $\ttbar$+W or $\ttbar$+Z, W+\Pgg~or Z+\Pgg, and diboson (ZZ, WZ, and WW)
processes. Single top quark events are generated with
the next-to-lowest-order (NLO) generator {\POWHEG 1.0}~\cite{Frixione:2007vw},
modeling the decay of $\tau$ leptons with
{\TAUOLA}~\cite{Was:2000st}. The Z2* tune~\cite{Field2,Field3} is used for
the underlying event. All simulated backgrounds are processed using the full
simulation of the response of the CMS detector using the
\GEANTfour~\cite{AGOSTINELLI2003250} package, and reconstructed under the
same conditions as the data. These  backgrounds are then normalized to the
integrated luminosity of the data using their respective cross
sections calculated at least at NLO. The CTEQ6M parton distribution functions
(PDF) are used in the signal and background simulations~\cite{Pumplin:2002vw}.
A summary of the software used in the MC simulations of backgrounds is given
in Table~\ref{table:softwarelist}.

\begin{table}[tbph]
  \centering
  \topcaption{Software used in MC simulations of backgrounds.}
\begin{tabular}{  l  c  c  }
  Software Package & Purpose & Physics Processes \\
    \hline
\MADGRAPH~5.1.3 & (LO) Matrix element generator & All backgrounds except single top \\
\POWHEG 1.0 & (NLO) Matrix element generator & Single top quark backgrounds \\
\PYTHIA~6.42 & Parton showering and fragmentation & All backgrounds and signal \\
{\TAUOLA} & Decay of $\tau$ leptons & Single top quark backgrounds \\
Z2* tune & Modeling of underlying event & All backgrounds and signal \\
\GEANTfour & Modeling of the CMS detector & All backgrounds and signal \\
CTEQ6M & Parton distribution functions (PDF) & All backgrounds and signal \\
\textsc{SuSpect} 2.41 & Generation of GGM signal spectrum & Signal \\
\PROSPINO 2.1 & (NLO) Cross section calculation & Signal \\
\SDECAY 1.2& Decay tables for GGM particles & Signal \\
    \hline
  \end{tabular}
  \label{table:softwarelist}
\end{table}

In the muon+jets channel, the background from Z+jets and Z+\Pgg~events is very small
because of the low probablility for a muon to be misidentified as a photon.
In the electron+jets channel, however, these processes contribute more to the background,
especially at low \ptmiss, because the probability for an electron to be misidentified as
a photon is much greater.
This electron misidentification rate can be  determined from the size of the peak at the Z
boson mass in the invariant mass distribution of
electron-photon pairs in the electron+jets channel
of SR1. This rate depends on an estimate of the number of selected Z bosons
in the electron+jets channel, the accuracy of which can be improved
through the implementation of a scale factor ($SF$) extracted
to normalize the Z+jets and Z+\Pgg~MC events in both the electron and muon channels.
The $SF$ is measured imposing a dilepton selection similar to the one used in the SR1 selection,
but altered to require two same-flavor leptons rather than just a single lepton.
Events with additional leptons are vetoed, and no photons
are required. A fit to the invariant mass of the dilepton
system in data, using the Z+jets and Z+\Pgg~MC events as the
signal template and all other MC events as background templates, provides a
normalization scale factor for both the
Z+jets and Z+\Pgg~MC events, labeled $SF_{\Z(\gamma)}$, in the electron and the muon channels.

\begin{figure}[tbph]
  \centering
    \includegraphics[width=0.6\textwidth]{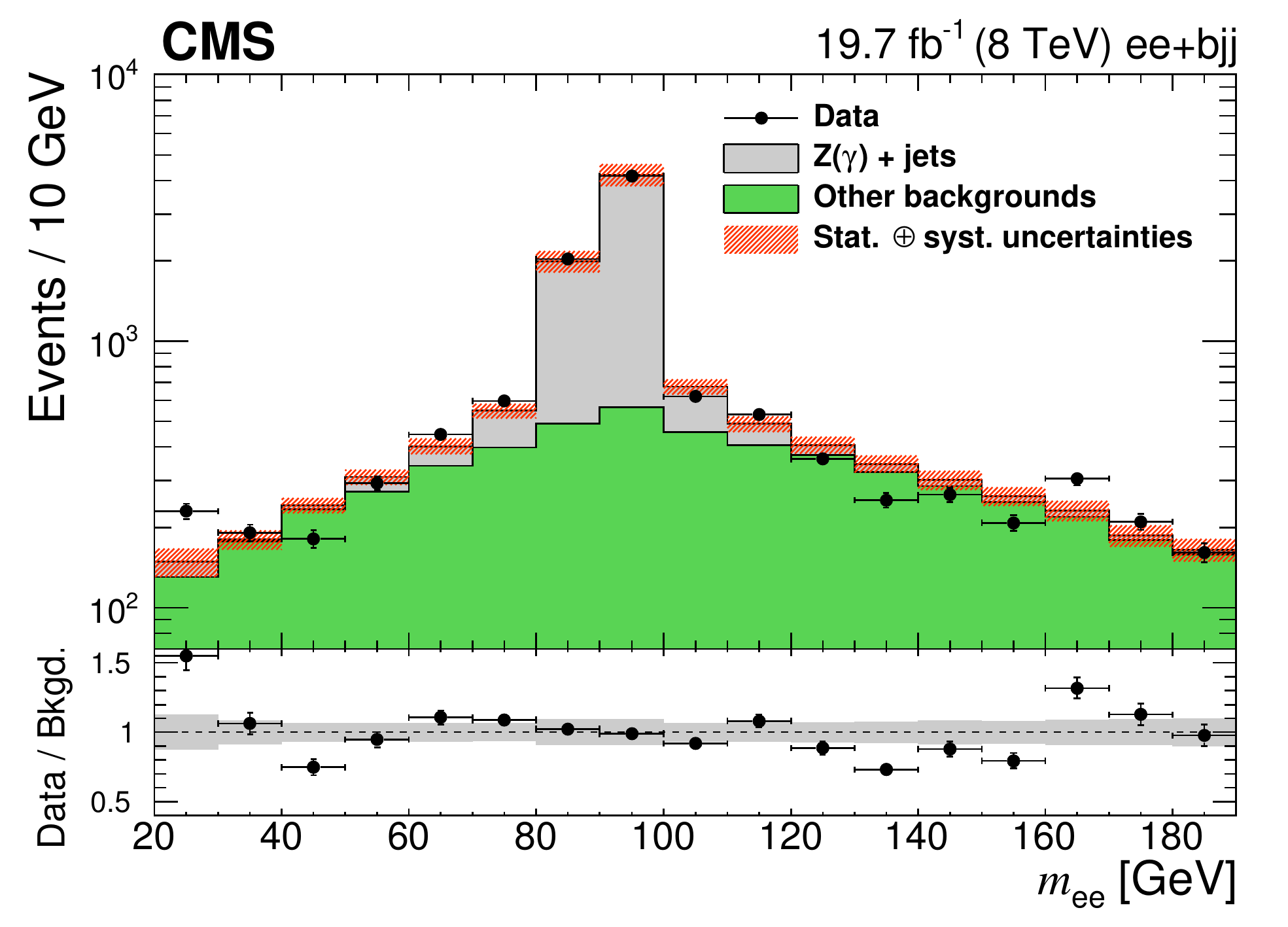}
    \includegraphics[width=0.6\textwidth]{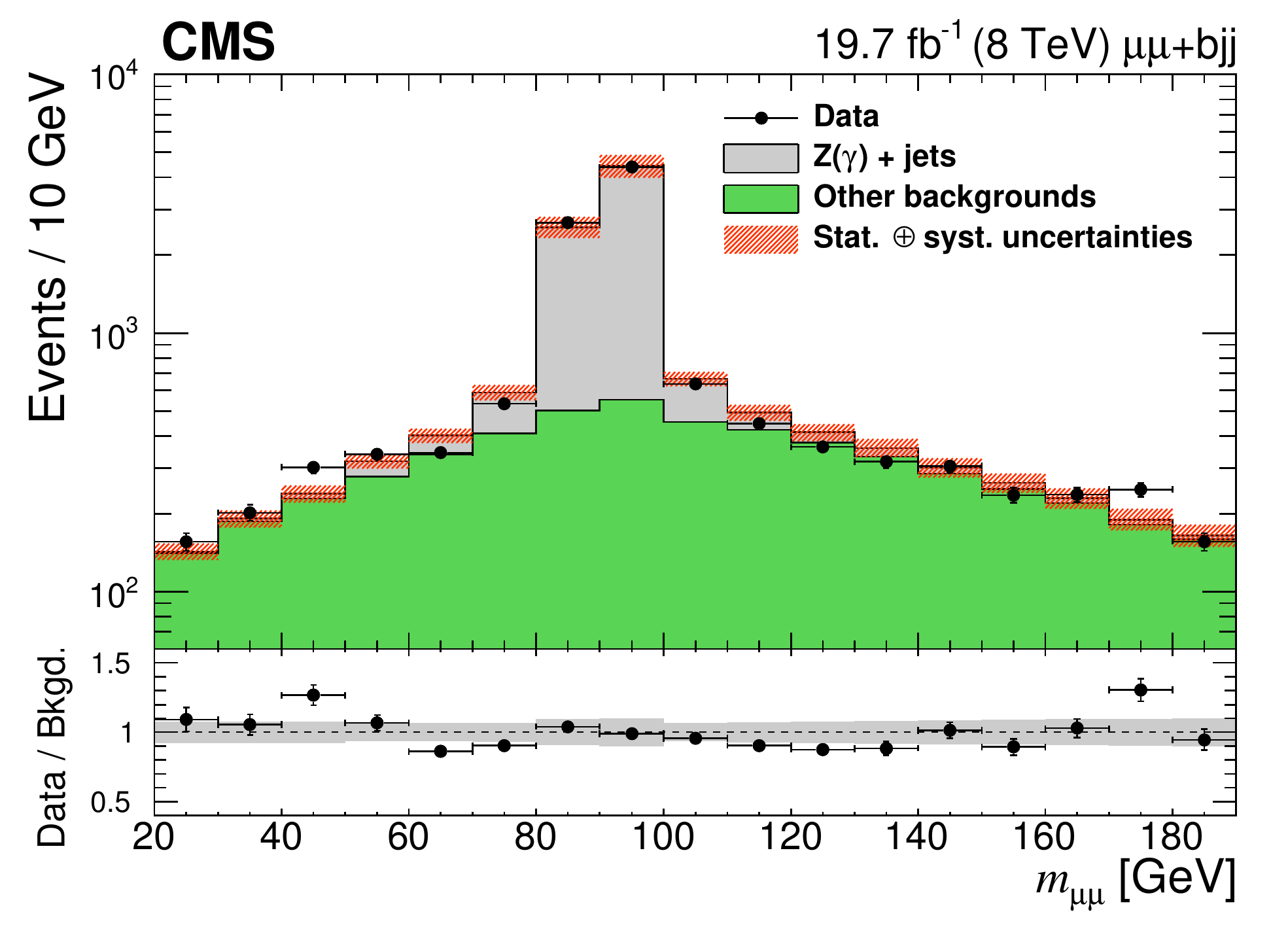}
    \includegraphics[width=0.6\textwidth]{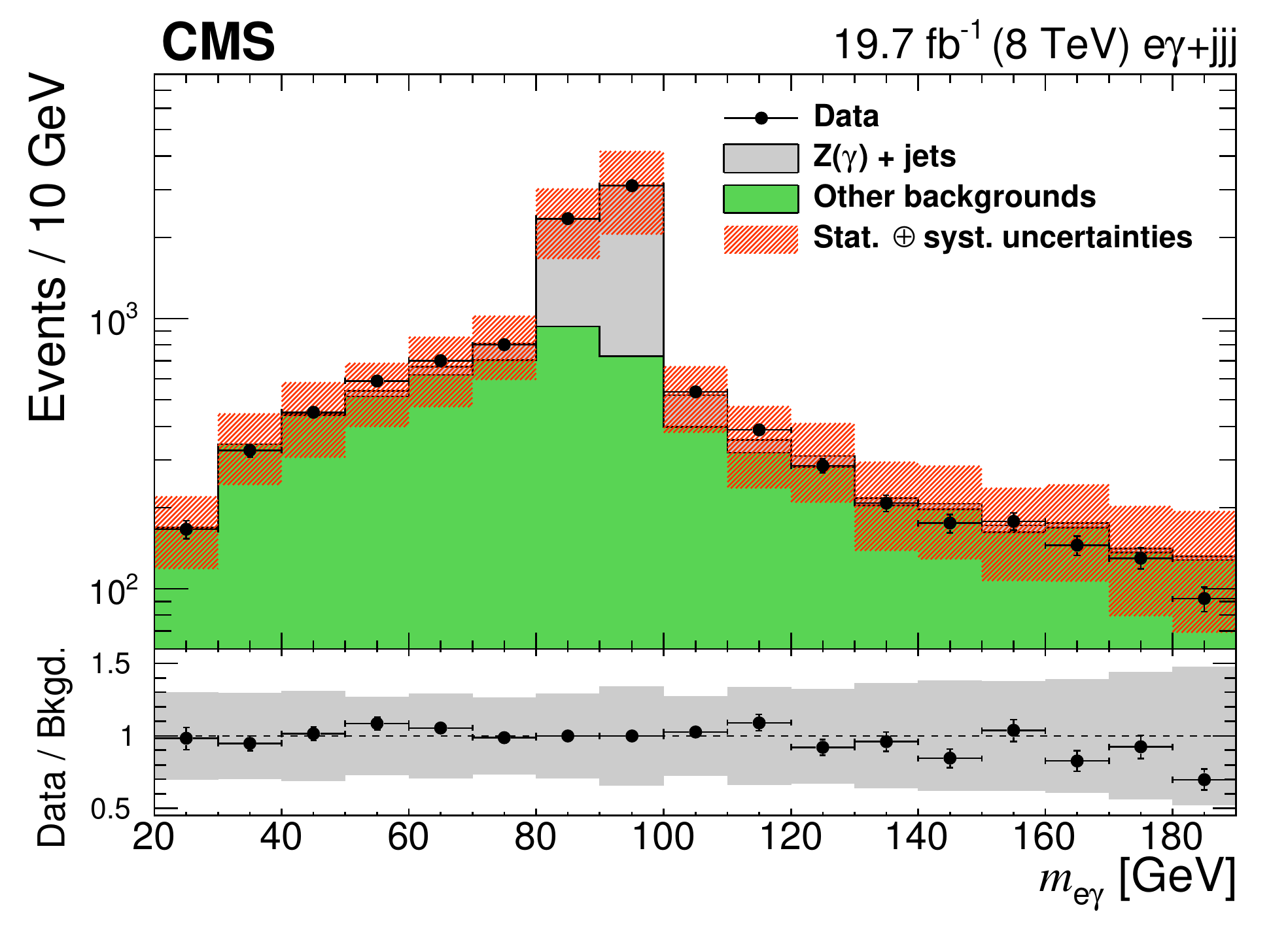}
    \caption{The dilepton invariant masses used in
      determining  $SF_{\Z(\gamma)}$ (upper pane) for the electron and (middle pane) for the muon
      channels.  The lower pane shows the result of the fit of $m_{\Pe\Pgg}$ SR1 electron data
      (without the b tag requirement) to determine
      $SF_{\Pe \to\gamma}$. The mass spectra are shown post-fit after the application of
      the derived scale factors.
      The ratio of data to the total background is included
      in the lower panel of each plot.  Uncertainties include the quadratic
      sum of all statistical and systematic components.}
    \label{fig:eleFakeRateResults}

\end{figure}

Once this first $SF$ is applied to correct the MC
estimate of the number of Z bosons, the Z resonance in the SR1 electron+jets
channel is used to obtain a second scale
factor $SF_{\Pe \to\Pgg}$ which corrects the misidentification of electrons as
photons. A fit to the invariant mass of the electron-photon
system in SR1 data, with $\ptmiss <50$\GeV, to limit the presence of signal,
is performed using the Z+jets and Z+\Pgg~MC events to determine their contributions.
Generator-level matching of reconstructed photons to
generated electrons is applied to increase the purity of the
misidentified e\Pgg~mass template. To increase the statistics available for each
template, the b tagging requirement is removed from the MC events and from the  data
sample, as the misidentification does not depend on the
presence of a b jet. From the result of this
fit, a normalization $SF_{\Pe \to\Pgg}$ is measured and applied to both the Z+jets and
Z+\Pgg~MC events in the electron-signal regions. A corresponding $SF_{\mu\to\Pgg}$ scale
factor is not applied in the muon-signal regions, as the
misidentification of muons as photons is negligible. The results
of the fits for each of these scale factors are listed
in Table~\ref{table:eleFakeRateScaleFactors}. Comparisons of the data and MC distributions are shown in
Fig.~\ref{fig:eleFakeRateResults} after the applying the scale factors of Table~\ref{table:eleFakeRateScaleFactors}.

\begin{table}[tbph]
  \centering
  \topcaption{Measured values of scale factors, $SF_{\Z(\gamma)}$ and $SF_{\Pe \to \gamma}$,
    used to correct the MC predictions for Z+jets and Z+\Pgg~backgrounds and electron-to-photon
    misidentification. For the electron+jets channel, the product
    of the two is applied to Z+jets and  Z+\Pgg~
    backgrounds. In the muon+jets channel, only the $SF_{\Z(\gamma)}$ scale factor is relevant.
    The first uncertainties are statistical, obtained from uncertainties in the resultant fits.
    The second uncertainties correspond to differences in the resulting scale factors,
    added in quadrature, that were obtained by allowing each systematic uncertainty to
    fluctuate up and down by one standard deviation and refitting.}
  \begin{tabular}{  l  c  c  }
    \hline
    Channel & $SF_{\Z(\gamma)}$ & $SF_{\Pe \to \gamma}$ \\
    \hline
    $\Pe$ & 1.38  $\pm$  0.02 $\pm$ 0.15 & 1.58 $\pm$ 0.03 $\pm$ 0.04 \\
    $\mu$ & 1.60  $\pm$  0.02 $\pm$ 0.17 & --- \\
    \hline
  \end{tabular}
  \label{table:eleFakeRateScaleFactors}
\end{table}

The final ingredient needed to estimate the background is
the relative compositions of photons and
photon-like jets in the dominant $\ttbar$+jets and $\ttbar$+\Pgg~backgrounds.
As stated in the introdcution, no
explicit $\ttbar$+\Pgg\Pgg~sample is used in the background estimate because of
the exceedingly small cross section for such events.
The sources of two photon events in SR2  are largely the result of jets
or electrons misidentified as photons as described above,
or of initial or final-state radiation as predicted by
\PYTHIA. While the precise photon
purity in each signal region is important for absolute
measurements, no difference in the overall shape of simulated \ptmiss  is
found when altering the purity of selected photons. The maximum
bin-by-bin difference between the simulated \ptmiss  of $\ttbar$+jets
and $\ttbar$+\Pgg~events is found to be 5\%. When their relative normalizations
are adjusted to the observed photon purity in data through a fit
to the  photon isolation variable, the result is well contained within the statistical uncertainties in the
\ptmiss  distribution. The \ptmiss distribution in both signal regions is
found to be insensitive to the source of selected photons in $\ttbar$+jets
and $\ttbar$+\Pgg~backgrounds, and, as such, no dedicated $\ttbar$+\Pgg\Pgg~sample
is required. To eliminate any dependence on the overall
production rate of $\ttbar$+\Pgg~events, the normalizations of $\ttbar$+jets and $\ttbar$+\Pgg~backgrounds
are allowed to float freely in the calculations of upper limits, so that
the interpretation of the results is based completely on the observed shapes of the distributions.

\begin{figure}[ptbh]
  \centering
    \includegraphics[width=0.7\textwidth]{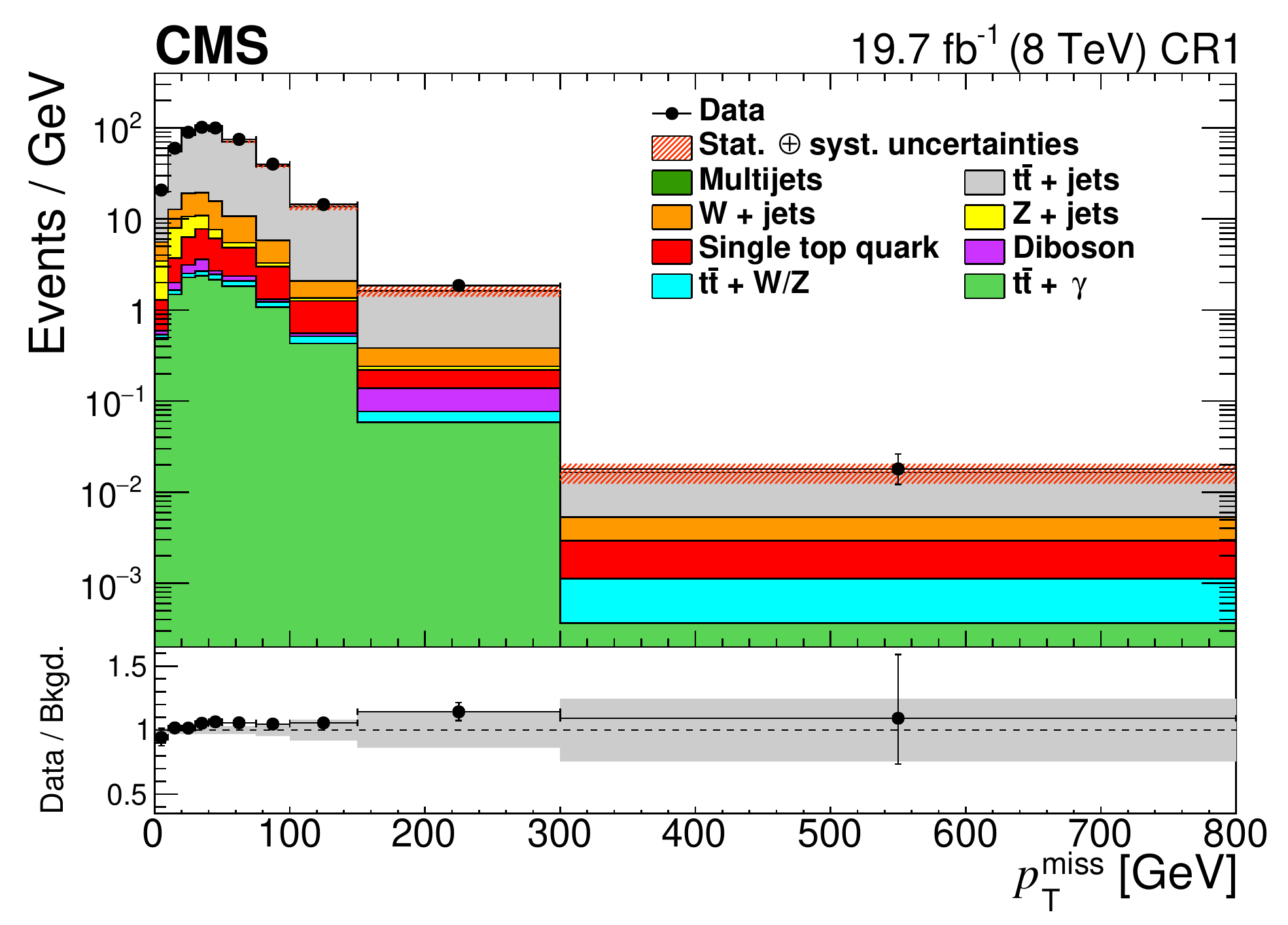}
    \includegraphics[width=0.7\textwidth]{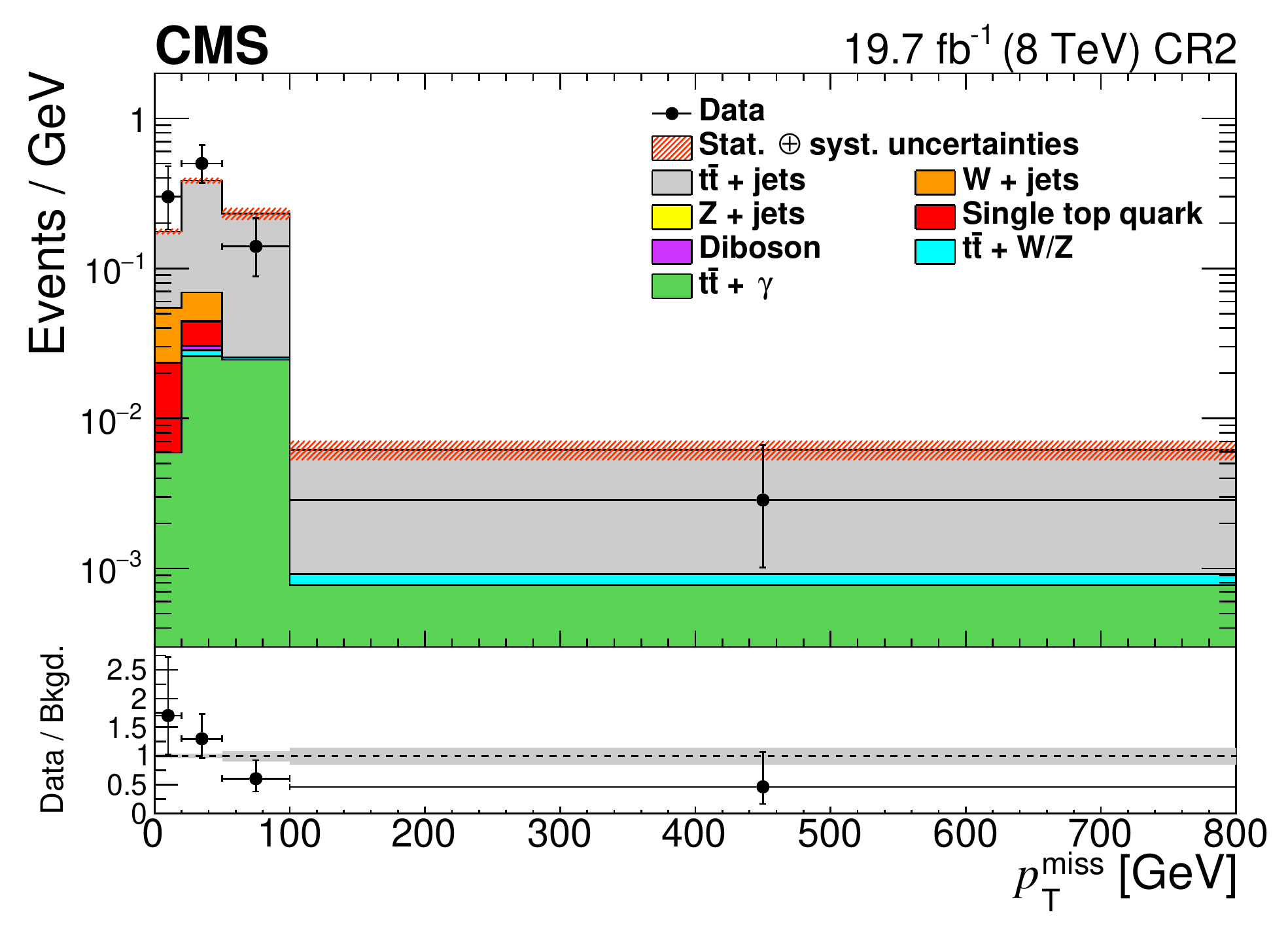}
    \caption{Comparison of data and simulated events as a function of \ptmiss for the combined
      e and $\mu$ control regions is shown: (upper pane) CR1 with one fake photon, and (lower pane) CR2
      with two fake photons. The content of each bin is normalized
      to its bin width.  The ratios of data to background are shown
      below the two panels. The overall uncertainties are obtained from the sum in
      quadrature of the statistical and systematic components. Note the Diboson background
      includes WW, WZ, ZZ, W+gamma, and Z+gamma.}
    \label{fig:controlRegionMET}

\end{figure}

The control regions allow us to validate the prediction of the \ptmiss
background, as they contain less than 1\% contamination from signal.
Inverting the requirements on the photon shower
selection or on charged-hadron isolation, the CR1 and CR2 regions can
contain the same \ttbar systems as the signal regions,
but with greatly enhanced contributions from  misidentified jets compared to the photon content in each sample.
The observed data and predicted background \ptmiss are shown in Fig.~\ref{fig:controlRegionMET}
for each control region.

The bin-by-bin fractional disagreement (1-Data/Background) ranging between 10-20\% between data and background in CR1
is taken as signal region systematic uncertainty in the modeling of \ptmiss, and is applied bin-by-bin
in the signal region. The Kolmogorov--Smirnov test~\cite{smirnov1948} result
of 0.66 between data and simulation for CR2
is attributable to the very small number of events in data and,
therefore, CR2 is not used to determine an  uncertainty for the signal region SR2.
The CR1 results are therefore used for both SR1 and SR2.
An additional systematic uncertainty in SR1 is obtained using the bin-by-bin fractional differences
($1-\mathrm{CR1/SR1}$) of CR1 and SR1 \ptmiss shapes.  A final systematic uncertainty
is obtained from a similar bin-by-bin difference ($1-\mathrm{SR1/SR2}$) for SR2.
Overall, this accounts for a 10--20\% systematic uncertainty
from differences between the data and the CR1 MC \ptmiss shapes, a 1--8\% systematic uncertainty in SR1 due to
the difference between CR1 and SR1 \ptmiss shapes, and a 10--50\%
systematic uncertainty (the 50\% value applies only in the highest bin of \ptmiss)
in SR2, based on the difference SR1 and SR2 \ptmiss shapes.

\section{Results and interpretation}
\label{interpretation}

For any given background or signal process, contributions from  systematic uncertainties affecting
\ptmiss are treated simultaneously and are assumed to be completely
uncorrelated. All backgrounds are simulated using MC generated events and assigned
systematic uncertainties based on integrated
luminosity uncertainties, PDF
and scale uncertainties, corrections for the number of pileup events, and
jet energy scale and resolution (JES and JER). Estimated uncertainties on trigger efficiency
and object selections are derived from the systematic
uncertainties in MC scale factors. These include trigger efficiencies,
b tagging~\cite{CMS-PAS-BTV-12-001,CMS-PAS-BTV-13-001} as well as
electron~\cite{cmsElectronReco}, muon~\cite{Chatrchyan:2012xi},
and photon identification~\cite{Khachatryan:2015iwa}.
The systematic uncertainties are summarized in Table~\ref{table:systematics}.

\begin{figure}[ptbh]
  \centering
    \includegraphics[width=0.7\textwidth]{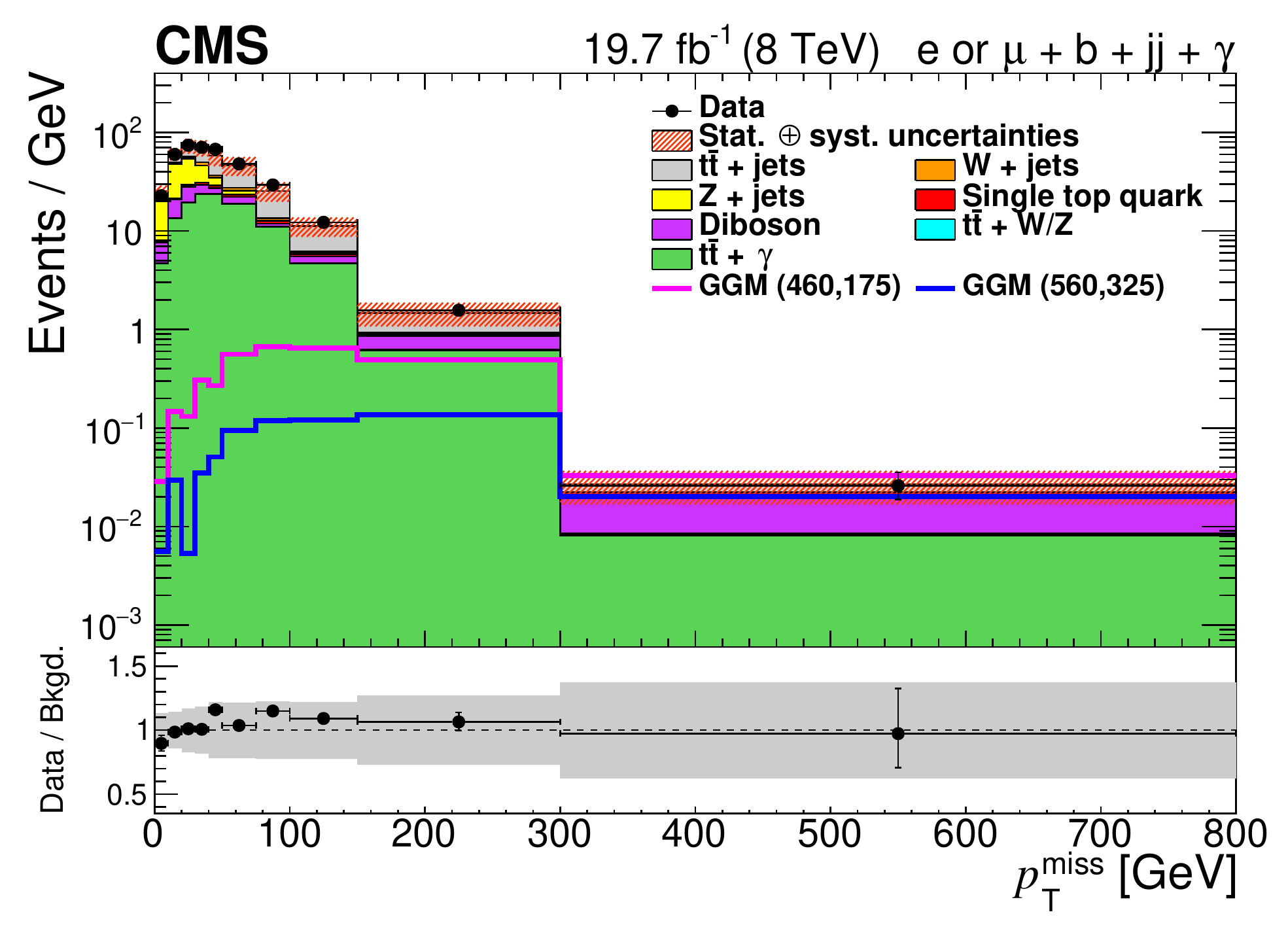}
    \includegraphics[width=0.7\textwidth]{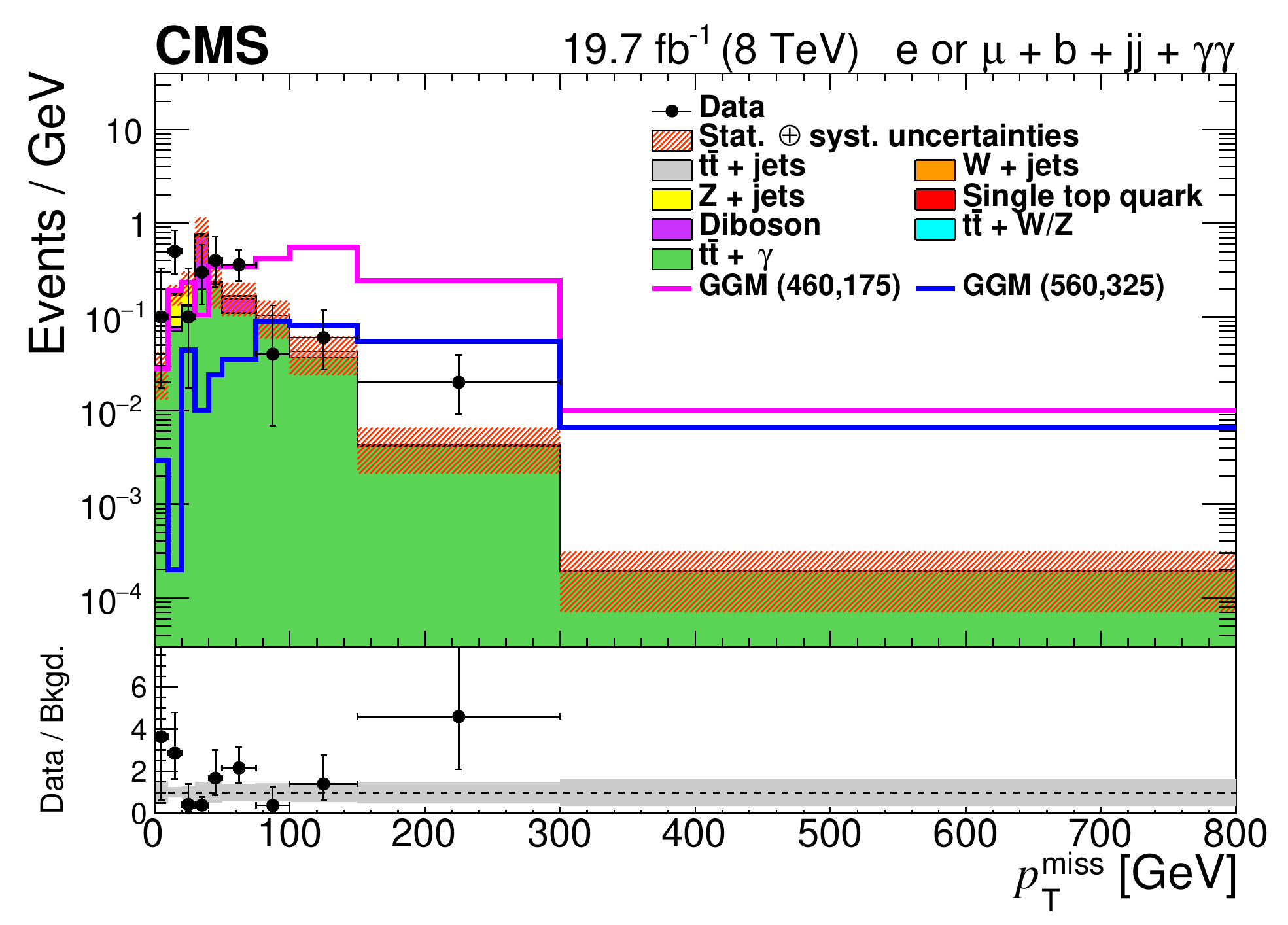}
    \caption{Comparison of data and MC simulation in \ptmiss for the combined
      (e and $\mu$) signal
      regions: (upper pane) SR1 with one reconstructed photon and (lower pane) SR2
      with two reconstructed photons. Each bin is normalized by its bin width.}
    \label{fig:metResults}

\end{figure}

The observed data are compared to the SM  background estimates as a function of
\ptmiss in each signal region, as shown in Fig.~\ref{fig:metResults}. No significant
deviation is observed between data and the background prediction.  The final results
are summarized in Table~\ref{table:finalResults}.

\begin{table}[tbph]
  \centering
  \topcaption{Summary of systematic uncertainties:  the
    dominant uncertainties are extracted from the control region. In the calculation of the upper limits,
    the normalizations of the $\ttbar$+jets and $\ttbar$+\Pgg~backgrounds are
    allowed to float freely in the fit. Check marks indicate the uncertainties that affect the
    shape of \ptmiss.}
    \resizebox{\textwidth}{!}{
  \begin{tabular}{  l  c  c  c  }
    \hline
    Source & Shapes & Uncertainty (\%)& Processes affected \\
    \hline
    Integrated luminosity & & 2.6 & Signal and  backgrounds \\
    Lepton ID/trigger & \checkmark & 1  & Signal and backgrounds \\
    Photon ID & \checkmark & 1.5  & Signal and backgrounds \\
    Pileup & \checkmark & 2  & Signal and backgrounds \\
    JES/JER & \checkmark & 5  & Signal and backgrounds \\
    b tagging & \checkmark & 2.5  & Signal and backgrounds \\
    \hline
    Renormalization and factorization scales & & 0.5 -- 25 & Backgrounds \\
    PDFs & & 2.5 -- 10  & Backgrounds \\
    \hline
    Control region discrepancy & \checkmark & 10 -- 20  & Backgrounds \\
    SR1/CR1 shape differences  & \checkmark & 1 -- 8  & Backgrounds \\
    SR2/SR1 shape differences & \checkmark & 10 -- 50  & Backgrounds \\
    \hline
    SUSY cross sections & & 16 -- 28  & Signal \\
    \hline
  \end{tabular}}
  \label{table:systematics}
\end{table}

\begin{table}[ptbh]
  \centering
  \topcaption{Observed data and expected background yields for the
    combined (e and $\mu$) signal regions. Expectations from two GGM signal model points are included,
    for which
    (460, 175) refers to $m_{\sTop} = 460\GeV$ and
    $m_{\chiz_1} = 175\GeV$, and similarly for (560, 325). The
    first group of uncertainties is statistical and the second is
    systematic.}
  \begin{tabular}{ lcc }
    \hline
    Channel & SR1 & SR2 \\
    \hline
    \ttbar + jets & 1845 $\pm$ 48 $\pm$ 64 & 1.42 $\pm$ 1.31 $\pm$ 0.12 \\
    W or Z + jets & 1100 $\pm$ 43 $\pm$ 35 & 2.12 $\pm$ 1.56 $\pm$ 0.22 \\
    Single t & 130 $\pm$ 14 $\pm$ 6 & --- \\
    Diboson & 22.7 $\pm$ 4.9 $\pm$ 1.1 & 0.20 $\pm$ 0.44 $\pm$ 0.06 \\
    V\Pgg~ & 431 $\pm$ 25 $\pm$ 116 & 6.2 $\pm$ 2.6 $\pm$ 3.8 \\
    $\ttbar$+W or Z & 14.7 $\pm$ 4.0 $\pm$ 1.0 & 0.15 $\pm$ 0.41 \\
    $\ttbar$+\Pgg~ & 1926 $\pm$ 47 $\pm$ 388 & 14.0 $\pm$ 4.0 $\pm$ 2.9 \\
    \hline
    Total background & 5469 $\pm$ 85 $\pm$ 411 & 24.1 $\pm$ 5.2 $\pm$ 4.8 \\
    \hline
    GGM (460, 175) & 162 $\pm$ 16 $\pm$ 6 & 87 $\pm$ 12 $\pm$ 2 \\
    GGM (560, 325) & 43.2 $\pm$ 7.5 $\pm$ 1.9 & 18.1 $\pm$ 4.8 $\pm$ 0.6 \\
    \hline
    Data & 5741 & 30 \\
    \hline
  \end{tabular}
  \label{table:finalResults}
\end{table}

To demonstated what a GGM signal would look like compared to the data, an example of
a GGM spectrum is generated with \textsc{FastSim}~\cite{1742-6596-331-3-032049}
using \PYTHIA 6 and \textsc{SuSpect} 2.41~\cite{Djouadi:2002ze}, using the decay tables from
\textsc{sdecay} 1.2~\cite{Muhlleitner:2004mka} and NLO cross sections
calculated with {\PROSPINO 2.1}~\cite{Beenakker:1996ed}. We scan over the
parameters $M_{1}$ ($U(1)_Y$), gaugino (bino) mass and $M_{tR}$ in the
SLHA files~\cite{Allanach:2008qq}. The other input parameters of GGM
such as $M_{2}$ ($SU(2)_L$), gaugino (wino) mass) and  $M_{dR}$, etc. are decoupled.
As a result {\SDECAY} + \textsc{SuSpect} produce neutralino and top squark masses that are similar
to the settings of $M_{1}$ and $M_{tR}$, and the rest of the particles masses are in the \TeV range.

The GGM signal is shown superimposed on the data and background MC
in Fig.~\ref{fig:metResults}. The mass of the top squark ($m_{\sTop}$) is chosen to range from 360 to 910\GeV. The neutralino
is assumed for simplicity to be 100\% bino-like, decaying 100\% to a photon plus a gravitino. The neutralino mass
($m_{\chiz_1}$) is chosen to range from 150 to 725\GeV and the gravitino mass is 1\GeV. Signal points are evaluated in 25\GeV
steps in both $m_{\chiz_1}$ and $m_{\sTop}$ up to 300\GeV, and in
50\GeV steps for higher masses.  All other SUSY particles (squarks, gluinos, and
gauginos) are decoupled by setting their masses to very large values so that the only relevant process is the
production of top squark pairs that decay to bino-like NLSPs. The mass region
where $m_{\sTop} - m_{\chiz_1} < m_{\PQt}$ is
not considered, as the requirement for high-\pt leptons and b jets
limits the sensitivity in this mass range.

No significant excess of events is observed beyond the SM expectation, and
95\% confidence level (CL) upper limits are placed on the  cross sections by
combining the results of all four search regions (electron SR1, muon SR1, electron SR2, and muon SR2)
using the CL$_\mathrm{s}$
criterion~\cite{CLS2,CLS1,Cowan}. The test statistic is constructed as the
product of likelihood ratios in bins of \ptmiss. Systematic uncertainties
are included as nuisance parameters in the signal and background \ptmiss shapes. Systematic
uncertainties affecting only the normalization
of signal or background processes are modeled through
log-normal distributions, taken as
the probability density functions in their associated
nuisance parameters. Fluctuations in the shape of
\ptmiss distributions determine both upward and downward systematic uncertainties.

A single $\pm$100\% nuisance parameter is introduced with a log-uniform probability
density function for its normalization to allow the \ttbar and $\ttbar$+\Pgg~normalizations
to float freely in the upper-limit calculation.
Statistical uncertainties resulting from the limited number of MC events
are also included as nuisance parameters, as prescribed in Ref.~\cite{BarlowBeeston}.

\begin{figure}[tbph]
  \centering
    \includegraphics[width=0.7\textwidth]{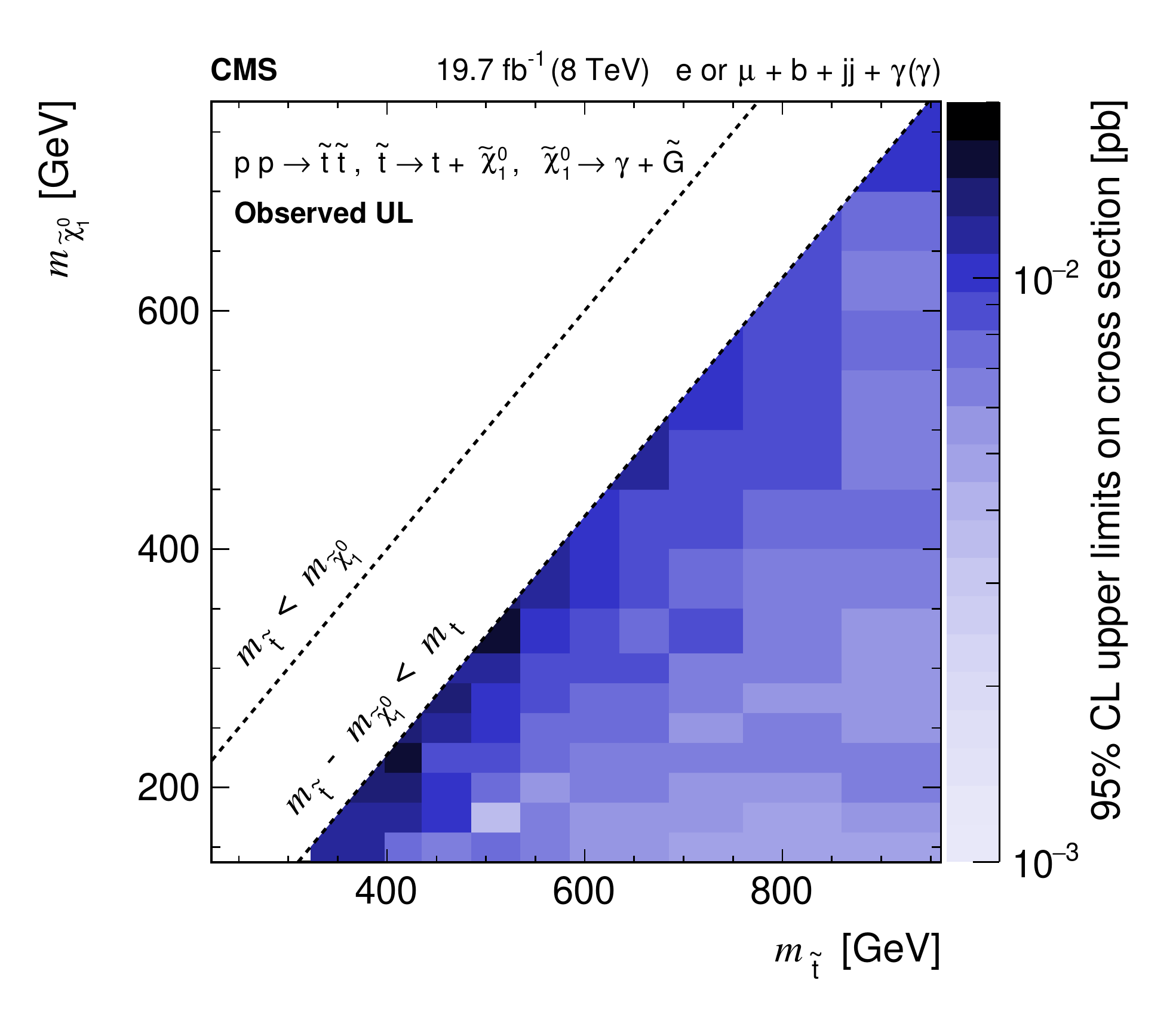}
    \includegraphics[width=0.7\textwidth]{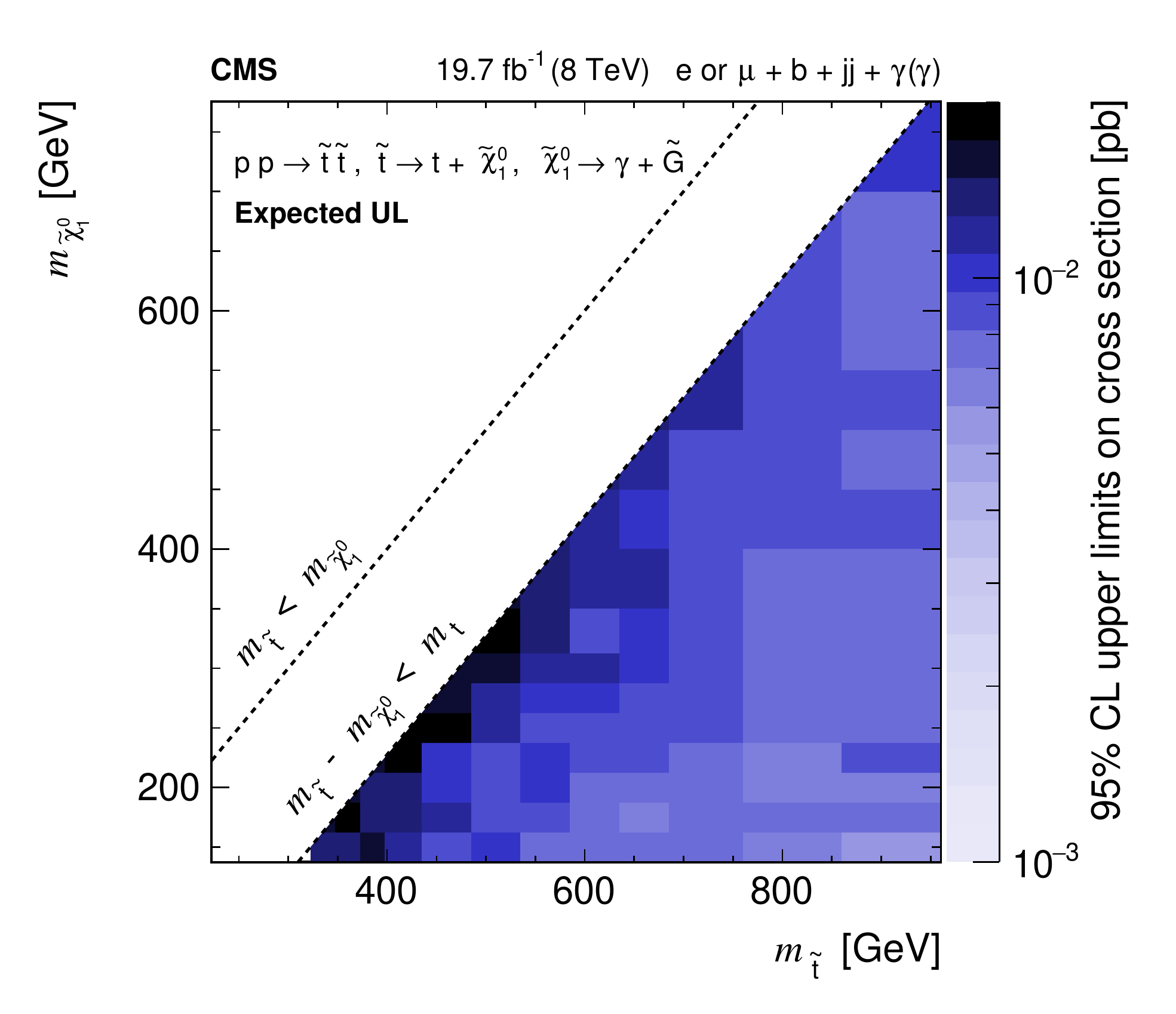}
    \caption{The observed (upper pane) and expected (lower pane) CL$_\mathrm{s}$ upper limits on the cross section  at 95\% CL in
     the $m_{\sTop}$ -- $m_{\chiz_1}$ plane.}
    \label{fig:upperLimits}

\end{figure}

\begin{figure}[tbph]
  \centering
    \includegraphics[width=0.7\textwidth]{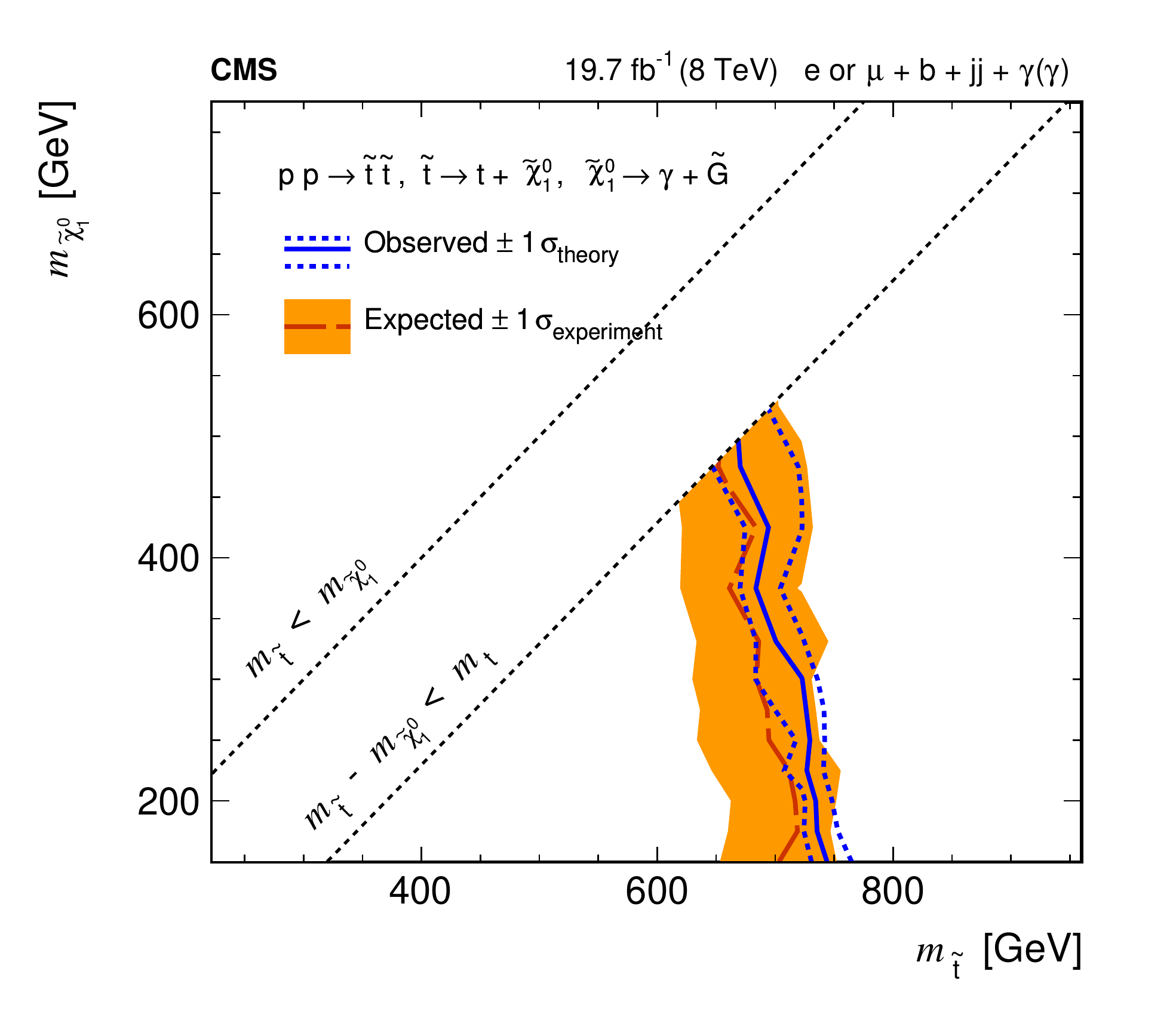}
    \caption{Observed and expected mean exclusions at the 95\% CL in the top squark and bino mass plane,
             and their ranges of uncertainties given by the contours at the 68\% CL. The region to the left of
             the contour for m$_{\sTop}$--m$_{\chiz_1}<$ m$_t$ is excluded by this analysis.}
    \label{fig:exclusionContour}

\end{figure}

The expected and observed upper limits are
shown in Fig.~\ref{fig:upperLimits}. The observed
upper limits are slightly less stringent than the expected limits. Observed and
expected exclusion contours are also determined and shown in
Fig.~\ref{fig:exclusionContour} with exclusion of top squark mass below
650 to 730\GeV corresponding to neutralino masses of 500 and 150\GeV, respectively.
These exclusions are obtained using the $-1 \sigma$ theoretical excursion
from the observed exclusion mean.

\section{Summary}
\label{Summary}

We have presented a search for natural gauge-mediated supersymmetry breaking
in events with a top quark pair and one or two photons. No significant
deviation is found in the distribution of the missing transverse momentum
between data and expected SM backgrounds that would
indicate the presence of new physics. Upper limits on signal cross sections are
calculated for a range of top squark and bino masses. Top squark masses
between 650 to 730\GeV are excluded at the 95\% CL corresponding to
the neutralino mass range of 500 to 150\GeV, respectively. These top squark mass points are obtained
using the $-1 \sigma$ theoretical excursion from the observed exclusion mean. These results
set the most stringent exclusions on top squark masses in gauge-mediated supersymmetric model
considered here.

\begin{acknowledgments}
\label{Acknow}
We congratulate our colleagues in the CERN accelerator departments for the excellent performance of the LHC and thank the technical and administrative staffs at CERN and at other CMS institutes for their contributions to the success of the CMS effort. In addition, we gratefully acknowledge the computing centers and personnel of the Worldwide LHC Computing Grid for delivering so effectively the computing infrastructure essential to our analyses. Finally, we acknowledge the enduring support for the construction and operation of the LHC and the CMS detector provided by the following funding agencies: BMWFW and FWF (Austria); FNRS and FWO (Belgium); CNPq, CAPES, FAPERJ, and FAPESP (Brazil); MES (Bulgaria); CERN; CAS, MoST, and NSFC (China); COLCIENCIAS (Colombia); MSES and CSF (Croatia); RPF (Cyprus); SENESCYT (Ecuador); MoER, ERC IUT, and ERDF (Estonia); Academy of Finland, MEC, and HIP (Finland); CEA and CNRS/IN2P3 (France); BMBF, DFG, and HGF (Germany); GSRT (Greece); OTKA and NIH (Hungary); DAE and DST (India); IPM (Iran); SFI (Ireland); INFN (Italy); MSIP and NRF (Republic of Korea); LAS (Lithuania); MOE and UM (Malaysia); BUAP, CINVESTAV, CONACYT, LNS, SEP, and UASLP-FAI (Mexico); MBIE (New Zealand); PAEC (Pakistan); MSHE and NSC (Poland); FCT (Portugal); JINR (Dubna); MON, RosAtom, RAS, RFBR and RAEP (Russia); MESTD (Serbia); SEIDI, CPAN, PCTI and FEDER (Spain); Swiss Funding Agencies (Switzerland); MST (Taipei); ThEPCenter, IPST, STAR, and NSTDA (Thailand); TUBITAK and TAEK (Turkey); NASU and SFFR (Ukraine); STFC (United Kingdom); DOE and NSF (USA).

\hyphenation{Rachada-pisek} Individuals have received support from the Marie-Curie program and the European Research Council and Horizon 2020 Grant, contract No. 675440 (European Union); the Leventis Foundation; the A. P. Sloan Foundation; the Alexander von Humboldt Foundation; the Belgian Federal Science Policy Office; the Fonds pour la Formation \`a la Recherche dans l'Industrie et dans l'Agriculture (FRIA-Belgium); the Agentschap voor Innovatie door Wetenschap en Technologie (IWT-Belgium); the Ministry of Education, Youth and Sports (MEYS) of the Czech Republic; the Council of Science and Industrial Research, India; the HOMING PLUS program of the Foundation for Polish Science, cofinanced from European Union, Regional Development Fund, the Mobility Plus program of the Ministry of Science and Higher Education, the National Science Center (Poland), contracts Harmonia 2014/14/M/ST2/00428, Opus 2014/13/B/ST2/02543, 2014/15/B/ST2/03998, and 2015/19/B/ST2/02861, Sonata-bis 2012/07/E/ST2/01406; the National Priorities Research Program by Qatar National Research Fund; the Programa Clar\'in-COFUND del Principado de Asturias; the Thalis and Aristeia programs cofinanced by EU-ESF and the Greek NSRF; the Rachadapisek Sompot Fund for Postdoctoral Fellowship, Chulalongkorn University and the Chulalongkorn Academic into Its 2nd Century Project Advancement Project (Thailand); and the Welch Foundation, contract C-1845.
\end{acknowledgments}
\bibliography{auto_generated}

\cleardoublepage \appendix\section{The CMS Collaboration \label{app:collab}}\begin{sloppypar}\hyphenpenalty=5000\widowpenalty=500\clubpenalty=5000\textbf{Yerevan Physics Institute,  Yerevan,  Armenia}\\*[0pt]
A.M.~Sirunyan, A.~Tumasyan
\vskip\cmsinstskip
\textbf{Institut f\"{u}r Hochenergiephysik,  Wien,  Austria}\\*[0pt]
W.~Adam, E.~Asilar, T.~Bergauer, J.~Brandstetter, E.~Brondolin, M.~Dragicevic, J.~Er\"{o}, M.~Flechl, M.~Friedl, R.~Fr\"{u}hwirth\cmsAuthorMark{1}, V.M.~Ghete, C.~Hartl, N.~H\"{o}rmann, J.~Hrubec, M.~Jeitler\cmsAuthorMark{1}, A.~K\"{o}nig, I.~Kr\"{a}tschmer, D.~Liko, T.~Matsushita, I.~Mikulec, D.~Rabady, N.~Rad, B.~Rahbaran, H.~Rohringer, J.~Schieck\cmsAuthorMark{1}, J.~Strauss, W.~Waltenberger, C.-E.~Wulz\cmsAuthorMark{1}
% ...
\vskip\cmsinstskip
% ...
1:~~Also at Vienna University of Technology, Vienna, Austria\\
% ...

\end{sloppypar}
\end{document}